\documentclass[12pt]{iopart}

\usepackage{graphicx}
\usepackage{xcolor}
\usepackage{gensymb}
\usepackage{changes}

\newcommand{\lSAW}{\lambda_\mathrm{SAW}}	
\newcommand{\fSAW}{f_\mathrm{SAW}}			
\newcommand{\TSAW}{T_\mathrm{SAW}}			

\newcommand{\Prf}{P_\mathrm{rf}} 			

\begin{document}
\title{Control of single photon emitters in semiconductor nanowires by surface acoustic waves}

\author{S. Lazi\'c$^1$, A. Hern\'andez-M\'inguez$^2$ and P. V. Santos$^2$}
\address{$^1$ Departamento de F\'isica de Materiales, Instituto Nicol\'as Cabrera and Instituto de F\'isica de Materia Condensada (IFIMAC), Universidad Aut\'onoma de Madrid, Francisco Tom\'as y Valiente 7, 28049 Madrid, Spain}
\address{$^2$ Paul-Drude-Institut f\"ur Festk\"orperelektronik, Hausvogteiplatz 5-7, 10117 Berlin, Germany}

\begin{abstract}
We report on experimental study into the effects of surface acoustic waves on the optical emission of dot-in-a-nanowire heterostructures in III-V material systems. Under direct optical excitation, the excitonic energy levels in III-nitride dot-in-a-nanowire heterostructures oscillate at the acoustic frequency, producing a characteristic splitting of the emission lines in the time-integrated photoluminescence spectra. This acoustically induced periodic tuning of the excitonic transition energies is combined with spectral detection filtering and employed as a tool to regulate the temporal output of anti-bunched photons emitted from these nanowire quantum dots. In addition, the acoustic transport of electrons and holes along a III-arsenide nanowire injects the electric charges into an ensemble of quantum dot-like recombination centers that are spatially separated from the optical excitation area. The acoustic population and depopulation mechanism determines the number of carrier recombination events taking place simultaneously in the ensemble, thus allowing a control of the anti-bunching degree of the emitted photons. The present results are relevant for the dynamic control of single photon emission in III-V semiconductor heterostructures. 
\end{abstract}


\submitto{\SST}


This Accepted Manuscript is available for reuse under a CC BY-NC-ND 3.0 licence after the 12 month embargo period provided that all the terms of the licence are adhered to.

\maketitle


\section{Introduction}

In the last years there has been an increasing interest in the use of quantum dots (QDs) embedded in semiconductor nanowires (NWs) as single photon sources (SPSs) for applications in quantum optics~\cite{Borgstroem_NL5_1439_05, Tribu_NL8_4326_08, Friedler_OE17_2095_09, Heinrich_APL96_211117_10, Reimer_NC3_737_12, Dalacu_NL12_5919_12}. There are several advantages of this configuration with respect to the use of QDs in conventional planar heterostuctures. First, the wetting layer is absent in the case of NWs. Carrier leakage from the QD into the wetting layer provides an escape pathway when operating at high temperatures, and thus degrades the photon emission intensity~\cite{Deshpande_APL102_161114_13}. Second, the inclusion of a single quantum emitter into a semiconductor NW strongly improves the emission efficiency and directionality, which are key requirements for long-distance quantum communications. Due to their specific geometry, semiconductor NWs guide the light from the embedded quantum emitter towards the extraction point at the top facets~\cite{Pelton_PRL89_233602_02, Friedler_OE17_2095_09, Claudon_NP4_174_10, Bleuse_PRL106_103601_11}, thus acting as natural photonic waveguides. In addition, tapering of the top region reduces the divergence of the output beam and hence further improves emission directionality~\cite{Reimer_NC3_737_12, Gregersen_OL33_1693_08}. Such highly directional emission is vital for efficient light collection into fibers or other optical pathways like integrated waveguides for e.g. quantum key distribution or on-chip quantum information processing. In this context, NWs offer a promising alternative to cavity-based solutions since they do not require tuning of the emitter into resonance with a cavity mode by post-growth manipulation. Third, NW heterostructures can be easily transferred from their native to a foreign substrate and, therefore, serve as fundamental building blocks for SPS integration into nanoscale optoelectronic and photonic devices~\cite{Borgstroem_NL5_1439_05, Huang_NL2_101_02, Qian_NL5_2287_05, Minot_NL7_367_07, Arafin_JoN7_74599_13}. Finally, an additional advantage of embedding QDs in NWs instead e.g. in quantum wells (QWs) is the small space provided by the NW geometry, which limits carrier diffusion and can consequently play an important role against the spectral diffusion of the emitted photons~\cite{Holmes_PRB92_115447_15}.

Among different materials available for the growth of dot-in-a-nanowire heterostructures, those based on (In, Ga)As or In(As, P) for photon emission in the near-infrared and on (In, Ga)N for the emission in the ultra-violet and visible spectrum are specially interesting. The large surface-to-volume ratio and small nanowire-substrate interface makes III-nitride NW heterostructures especially appealing, as it enables strain-free heteroepitaxy on largely mismatched substrates. The better strain relief reduces the formation of threading dislocations and other structural defects, resulting in high structural quality and improved optical performance (very weak absorption and emission below the band-edge), which is essential for obtaining long temporal and spatial single-photon coherence. Also, III-nitride QD systems are promising for technological applications as they can sustain room-temperature stable excitons. Finally, the negligible spectral contamination arising from the yellow band emission (typically present in other GaN-based systems) allows the production of single photons in a wide range of the visible spectrum. 

For NWs emitting in the infrared, the vapor-liquid-solid mechanism commonly used to grow GaAs and InP NWs typically generates structures with alternating sections of zincblende and wurzite crystal phases \cite{Caroff_IJoSTiQE17_829_11}. It has been shown that the combination of these two polytypes in a NW can lead to SPSs based on crystal phase QDs \cite{Akopian_NL10_1198_10}. However, the presence of charge traps at zincblende-wurzite interfaces close to a quantum dot can also enhance spectral diffusion \cite{Sallen_PRB84_41405_11}, thus degrading the purity of the excitonic emission required for quantum optic applications. Improvements in the growth techniques have demonstrated, e.g., ultraclean emission of In(As, P) quantum dots embedded in defect-free wurzite InP NWs \cite{Dalacu_NL12_5919_12}.

To fully exploit all these favorable features, appropriate techniques for the control of charge carriers in semiconductor NWs are needed. Although this can be easily achieved by using electric fields, local carrier generation in a well-defined area of the NW typically requires selective doping of NW regions, as well as the deposition of electric contacts on structures with a non-planar geometry~\cite{Haraguchi_APL60_745_92, Minot_NL7_367_07}. These challenges can be overcome by the remote carrier control provided by surface acoustic waves (SAWs). SAWs are elastic vibrations in a solid that propagate confined to its surface~\cite{Auld90a, Royer00a}. If the material is piezoelectric, then the vibrations are accompanied by an oscillating electric field, which extends typically a few micrometers both below and above the surface \cite{Lewis_37_85}. It has been shown that SAWs with wavelengths in the micrometer range are a powerful tool to confine and transport free carriers~\cite{Rocke97a}, excitons~\cite{PVS260}, spin ensembles~\cite{PVS152}, and even to modulate exciton-polariton condensates in 2D semiconductor heterostructures below the surface~\cite{PVS223}. In addition, SAWs have successfully been used to control optically active QDs embedded in these planar structures \cite{Wiele98a, Gell_APL93_081115_08, PVS218, Pustiowski2015}.

SAWs have also been employed in hybrid systems consisting of a semiconductor structure placed on an piezoelectric insulating substrate for the transport of electrically injected carriers in carbon nanotubes~\cite{Ebbecke_PRB70_233401_04, Leek_PRL95_256802_05, Buitelaar_PRL101_126803_08} and in graphene layers \cite{Miseikis_APL100_133105_12, PVS269, Bandhu_APL103_133101_13}. In the case of semiconductor NWs, this approach has been applied to demonstrate not only acousto-electric currents~\cite{Ebbecke_N19_275708_08}, but also to locally modulate the photon emission properties of both the NW core \cite{Kinzel_NL11_1512_11} and of QDs embedded in the NW \cite{Weiss2014, Weiss_NL14_2256_14, Lazic_AIPAdv_2015, Kinzel2016}. Finally, SAW-induced transport of photo-excited electrons and holes along the axial direction of a single NW has been demonstrated \cite{PVS246, PVS273}. These results open new perspectives for the fabrication of SPSs based on the acoustic pumping of electric charges into QDs contained in semiconductor NWs. Such sources combine the high photon extraction efficiency enabled by the NW geometry with the low jitter levels and high repetition rates arising from the acoustic pumping.

In this contribution, we review our experiments on the dynamic control of the photo-emission from QD-like recombination centers embedded in III-V semiconductor NWs by SAW-induced strain and piezoelectric fields. The examples discussed here are based on two types of NWs both containing Ga and In as III-components, but a different V-element in each case. In the first one, we show that SAWs can be used to modify the photon emission dynamics of SPSs located at the top of a GaN NW. The effect is based on the acousto-mechanical coupling of the propagating SAW to the nanowire, resulting in a periodic modulation of the QD transition energies. This, combined with spectral filtering, enables the photon output to be clocked at the acoustic frequency~\cite{Gell_APL93_081115_08}. In the second one, the SAW-induced transport of photo-excited electrons and holes in a GaAs NW into recombination centers placed at its edge acts as a source of anti-bunched photons~\cite{PVS246}. We show that the degree of antibunching depends strongly on the SAW intensity, thus demonstrating that the number of photons simultaneously emitted is controlled by the acoustic mechanism of population and depopulation of the recombination centers.

\section{Methods}

\subsection{Growth of nanowire heterostructures}

Ordered GaN nanowire arrays hosting (In,Ga)N nano-disks were fabricated using a bottom-up approach by selective area growth (SAG) plasma-assisted by molecular beam epitaxy (PA-MBE) on (0001) GaN-on-sapphire templates covered by titanium nanohole masks deposited by colloidal lithography~\cite{Gacevic_Book_2016}. The nanowires have a typical height of $\sim$500~nm and a diameter of $\sim$200~nm. They exhibit hexagonal cross section with a truncated pyramidal top. This pencil-like morphology determines the shape of the (In,Ga)N nano-disks embedded inside the GaN nanowire tips. The (In,Ga)N nano-disks are grown with varying thickness and In compositions (up to 30~nm and $\sim$20$\%$, respectively, depending on the disk region) and are capped with a 50~nm thick GaN~\cite{Gacevic_PRB_93_125436}. As detailed in Refs.~\cite{Chernysheva2015, ProcSPIE_9363}, the QD-like emission centers under study are formed by fluctuations of the In content in the topmost (In,Ga)N disk region. 

The GaAs NWs were grown by molecular beam epitaxy (MBE) using a self-assisted vapor-liquid-solid growth process on a Si(111) substrate~\cite{Breuer_NL11_1276_11}. Although SAW control has also been reported in radially grown GaAs QDs~\cite{Weiss2014}, this approach leads to recombination centers that are randomly distributed along the NW axis~\cite{Heiss_PRB83_45303_11}. In our experiment, the (In,Ga)As recombination centers were formed at the end of the axial growth, and therefore they are confined at one of the NW edges. First, an undoped GaAs core was grown with a diameter of 106$\pm$18~nm and several micrometers in length. During the last steps of the core deposition, QD-like recombination centers were incorporated by exposing the growing surfaces to In and As fluxes. This lead to the formation of an approximately 300~nm long segment at the top of the NWs containing In inclusions.  Due to the modified MBE growth conditions, the (In,Ga)As segment has a larger diameter (200~nm) and a higher density of planar defects than the GaAs section, as confirmed by transmission electron microscopy (not shown here). Next, the NWs were coated with a shell consisting of 22$\pm$9~nm AlAs and a thin GaAs layer. The final average length of the NWs is 7~$\mu$m.

\subsection{Characterization of SAW delay line}

\begin{figure}
\begin{center}
\includegraphics[width=.5\linewidth]{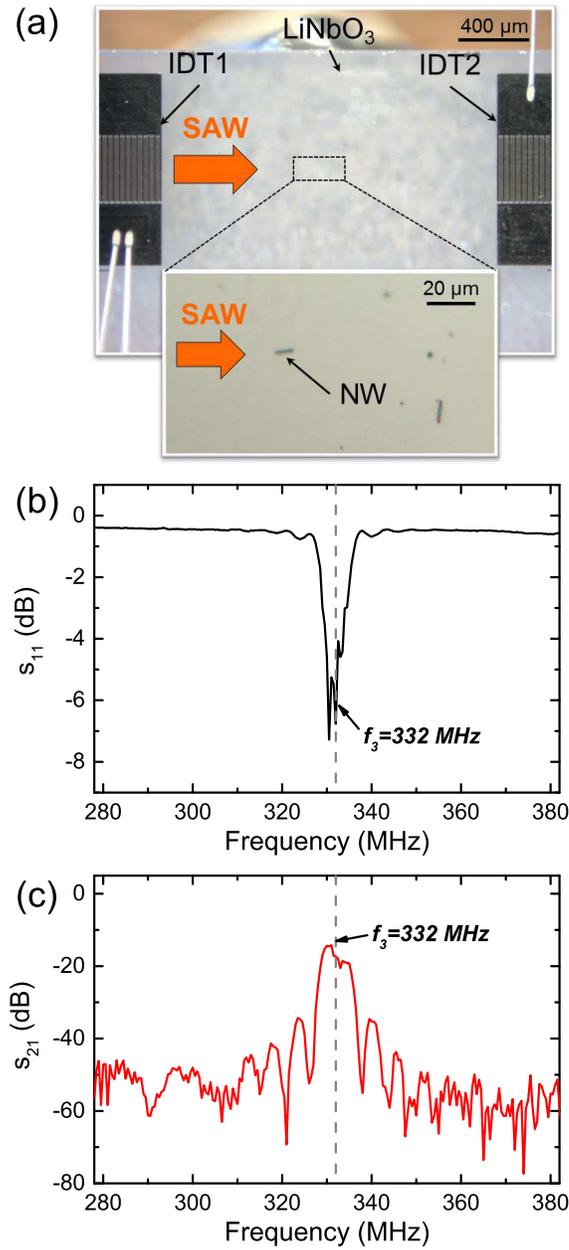}
\caption{(a) Acoustic delay line consisting of two interdigital transducers (IDTs) photo-lithographically patterned on the surface of a LiNbO$_3$ crystal. The semiconductor nanowires (NWs) are mechanically transferred onto the area between the IDTs. (b) Dependence of the rf power reflection coefficient, $s_{11}$, on the frequency of the rf signal applied to the IDT$_1$. The plot displays the region around the resonant frequency used in our experiments, $f_3=332$~MHz. (c) Corresponding rf power transmission coefficient between IDT$_1$ and IDT$_2$, $s_{21}$, measured in the same frequency range as in panel (b).}
\label{fig_setup}
\end{center}
\end{figure}

Both GaN/(In,Ga)N and GaAs/(In,Ga)As nanowires were mechanically transferred to acoustic delay lines. The delay line consists of two interdigital transducers (IDTs) deposited by optical lithography on the surface of the 128$\degree$~Y-cut of a LiNbO$_3$ crystal see Fig.~\ref{fig_setup}(a). As LiNbO$_3$ is a strongly piezoelectric material, a radio-frequency (rf) signal of the appropriate frequency applied to IDT$_1$ is transformed into a SAW by means of the inverse piezoelectric effect. During the SAW propagation along the delay line, the strain field is accompanied by an electric field due to the direct piezoelectric effect, which transforms the SAW back into an rf signal when it reaches IDT$_2$. Both the strain and piezoelectric fields can modulate the electronic band structure of the semiconductor NWs deposited on its surface.

The IDTs constituting our acoustic delay lines are floating electrode unidirectional transducers~\cite{Yamanouchi92a} designed to generate multiple harmonics of a fundamental wavelength $\lambda_1=35~\mu$m. The experiments reported here were carried out using the third harmonic with $\lambda_3=11.67~\mu$m. It corresponds to a resonance band centered at frequency $f_3=332$~MHz at room temperature. Figs.~\ref{fig_setup}(b) and \ref{fig_setup}(c) show the rf power reflection and transmission coefficients, $s_{11}$ and $s_{21}$ respectively, around the resonance. They were carried out with an rf connection configuration like the one used in the low temperature measurements. The frequency bandwidth (defined as the frequency range where the transmission is within 3~dB of its maximum value) of the IDTs is 3.5~MHz, and the insertion loss of the delay line (i.e.~the $s_{21}$ maximum at the resonance center) is -14.6~dB for $f_3$. Since the IDTs at each side of the delay line are equal, the electro-acoustic conversion efficiency is simply half of the insertion loss. Therefore, for the third harmonic, the IDTs convert approximately 20\% of the applied rf power into acoustic power.

\subsection{Photoluminescence and photon autocorrelation measurements}

The SAW delay line was mounted in a cold finger cryostat equipped with an optical window for spatially resolved micro-photoluminescence ($\mu$-PL) measurements and with coaxial connections for the application of rf signals to the IDTs. The measurements were carried out on NWs with the growth axis almost parallel to the SAW propagation direction, cf. Fig.~\ref{fig_setup}(a). The excitation source for the GaN/(In,Ga)N NWs was a 442~nm continuous wave (cw) laser (the corresponding excitation energy was therefore below the GaN bandgap), while a 757~nm pulsed diode laser was used in the case of the GaAs/(In,Ga)As NWs. The laser beam was focused onto a single NW by a microscope objective with high magnification power (diameter of the laser spot $\sim{1.5}~\mu$m). The PL emitted by the NW was collected by the same objective and imaged on the entrance slit of a single-grating spectrometer. Time-integrated detection with $\sim{200}$~nm spatial and $\sim{350}~\mu$eV spectral resolution was performed by a liquid N$_2$ cooled charge-coupled device (CCD) camera. The polarization properties of the GaN/(In,Ga)N dot-in-a-nanowire emission were analyzed using a half-wave retardation plate and a fixed linear polarizer placed in the collection path.

For the investigation of the photon emission statistics, we first spectrally filtered the PL emitted by the QD-like centers to select an isolated emission line. This was then analyzed using a Hanbury Brown and Twiss (HBT) setup consisting of two Si avalanche photodiodes (APDs), each placed at one of the two output ports of a 50/50 non-polarizing beam-splitter \cite{Lounis_RoPiP68_1129_05}. The photodetectors were connected to either the ``start'' or ``stop'' inputs of a time-correlated single photon counting (TCSPC) electronics, which delivers a histogram of the temporal distribution of coincidences between the two photodetectors. When triggered by the laser pulses, the same setup also yields time-resolved PL traces.

\section{Experimental Results}

\subsection{SAW modulation of single photon emitters based on GaN/(In,Ga)N nanowires}

\begin{figure}
\begin{center}
\includegraphics[width=.8\linewidth]{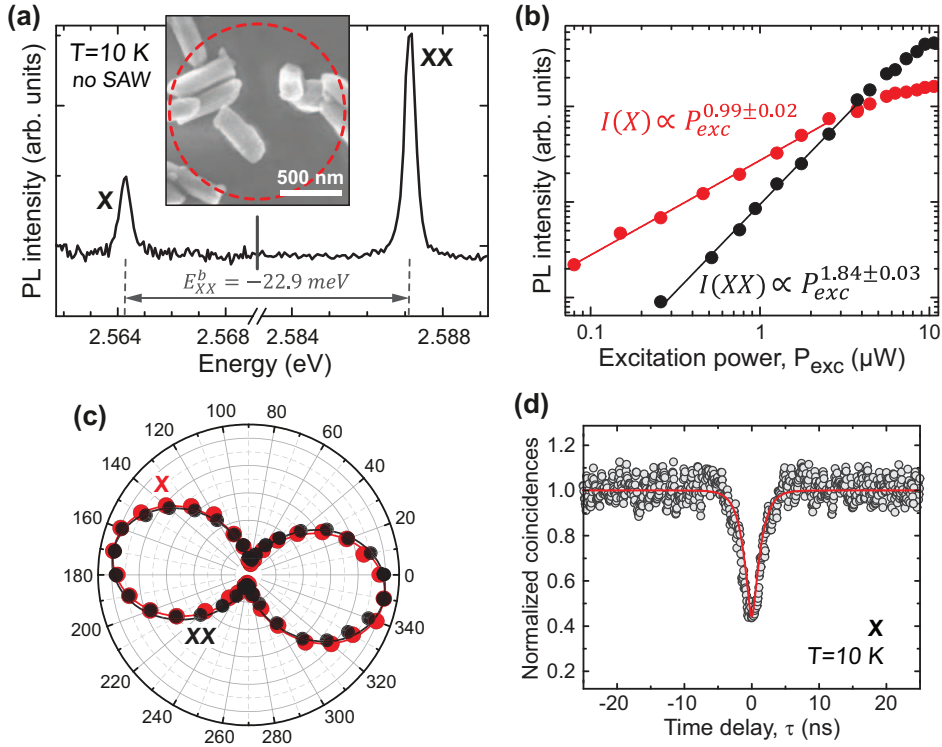}
\caption{(a) Low-temperature $\mu$-PL spectrum of the selected GaN/(In,Ga)N nanowire quantum dot showing the exciton (X) and biexciton (XX) emission lines in the absence of the acoustic excitation. Inset: A SEM top view of GaN/(In,Ga)N NW heterostructures mechanically transferred onto the SAW chip. The NWs are optically excited by the focused laser spot marked with dashed circle. (b) Integrated PL peak intensities of the X (red circles) and XX (black circles) as a function of the optical pump power. Lines are power-law fits to the experimental data. (c) Angular polarization distribution of the X (red circles) and XX (black circles) emission intensity with respect to a randomly chosen horizontal axis. (d) Autocorrelation histograms of the X and XX emission peaks.}
\label{fig_S1}
\end{center}
\end{figure}

The photoluminescence of (In,Ga)N QDs embedded into GaN NWs dispersed on the SAW propagation path was probed under direct illumination. As determined from the scanning electron microscopy (SEM) image of the probed sample region, cf. inset in Fig.~\ref{fig_S1}(a), more than one NW (typically up to 10) were optically excited by the focused laser spot. Moreover, only a small fraction of randomly dispersed nanowires showed efficient coupling to the SAW. In fact, the strongest coupling was observed in nanowires whose pyramidal tip (containing the QD-like emission centers) was in direct mechanical contact with the underlying LiNbO$_3$ substrate \cite{Lazic_AIPAdv_2015}.

Figure~\ref{fig_S1}(a) displays the low-temperature $\mu$-PL spectrum recorded at T=10~K on one of the probed nanowire QDs exhibiting strong electromechanical coupling to the SAW delay line. In the absence of the SAW the spectrum shows narrow PL peaks corresponding to recombination of exciton (X) and biexciton (XX) complexes. The X and XX assignment of the observed emission lines is corroborated by excitation-power dependence measurements, which present linear and nearly quadratic increase, respectively, of the integrated peak intensities with increasing optical pump power, cf. Fig.~\ref{fig_S1}(b). Additional confirmation is provided by the polarization-resolved PL measurements~\cite{Amloy_APL99_251903_11}, which show co-linear polarization of the two PL peaks with a similar polarization degree of $\sim{82}\%$, cf. Fig.~\ref{fig_S1}(c). Such high degree of linear polarization is ubiquitous in III-nitride-based QDs and can be ascribed to the valence band mixing induced by the in-plane anisotropy of the QD shape, as well as the anisotropy of the internal strain and electric fields~\cite{Bardoux_PRB77_235315_08, Winkelnkemper_JAP101_113708_07}. Note also that the negative binding energy (E$_{XX}^b$=E$_{X}$-E$_{XX}$=-22.9~meV), commonly found in c-plane III-nitride QDs with inherently large built-in electric fields~\cite{Simeonov_PRB77_75306_08}, is indicative of a strong localizing potential counteracting the Coulomb repulsion within the four-particle biexciton complex~\cite{Sebald_pssc6_872_09, Amloy_N25_495702_14}.

The autocorrelation histogram of the X emission recorded at $\sim{10}$~K under cw optical excitation is presented in Fig.~\ref{fig_S1}(d). The histogram reveals a pronounced anti-bunching dip at zero-time delay ($\tau=0$) with the anti-bunching value of $g^{(2)}(0)=0.07\pm{0.04}$ obtained after correction for finite temporal resolution of the HBT setup and the contribution of background photons (including the APDs’ dark counts and the contribution from additional emission centers within the detection area)\cite{Chernysheva2015, ProcSPIE_9363}. This value is well below the 0.5 two-photon threshold, thus proving the quantum nature of light emitted from the probed GaN/(In,Ga)N dot-in-a-nanowire.

\begin{figure}
\begin{center}
\includegraphics[width=.8\linewidth]{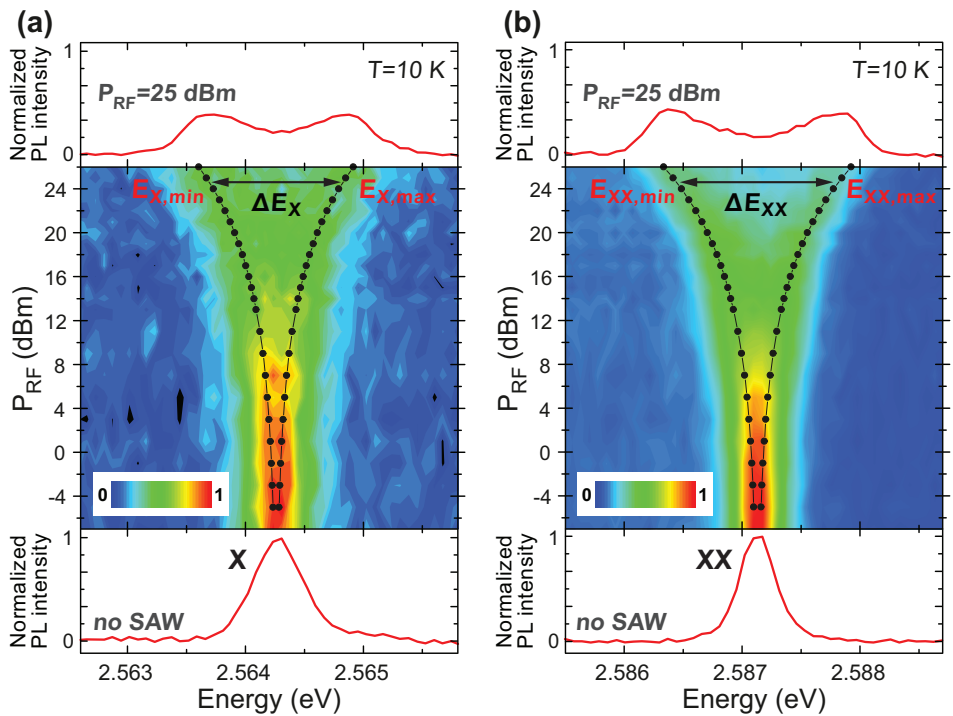}
\caption{False-color plot showing splitting of the QD X (a) and XX (b) PL lines as a function of the SAW power, $\Prf$. Bottom and top panels display the unperturbed and SAW-broadened X and XX emission spectra, respectively. }
\label{fig_S2}
\end{center}
\end{figure}

When the SAW is applied (with a frequency $f_3=338$~MHz at low temperature) the emission spectrum undergoes an apparent splitting of the two emission lines (Fig.~\ref{fig_S2}(a)~and~\ref{fig_S2}(b), top panels). The observed spectral changes can be mainly attributed to the dynamic modulation of the QD energy levels induced by the deformation potential associated with the periodic SAW strain field~\cite{Lazic_AIPAdv_2015}. This, in turn, causes the QD transition energies to oscillate around their equilibrium value in the absence of a SAW. In this context, the energy separation between the split doublet lines ($\Delta{E}_{X(XX)}$) reflects the difference between the transition energies at SAW phases corresponding to maximum compressive ($E_{X(XX),max}$) and tensile ($E_{X(XX),min}$) strain. The magnitude of the line splitting (i.e. $\Delta{E}_{X(XX)}$=$E_{X(XX),max}$-$E_{X(XX),min}$) increases continuously with increasing the RF power applied to the IDT (Fig.~\ref{fig_S2}, middle panels). These findings indicate that the SAW can be used to fine-tune the QD emission energy within a bandwidth up to $\sim$1.5~meV for the highest acoustic power $\Prf$=25~dBm.

\begin{figure}
\begin{center}
\includegraphics[width=.65\linewidth]{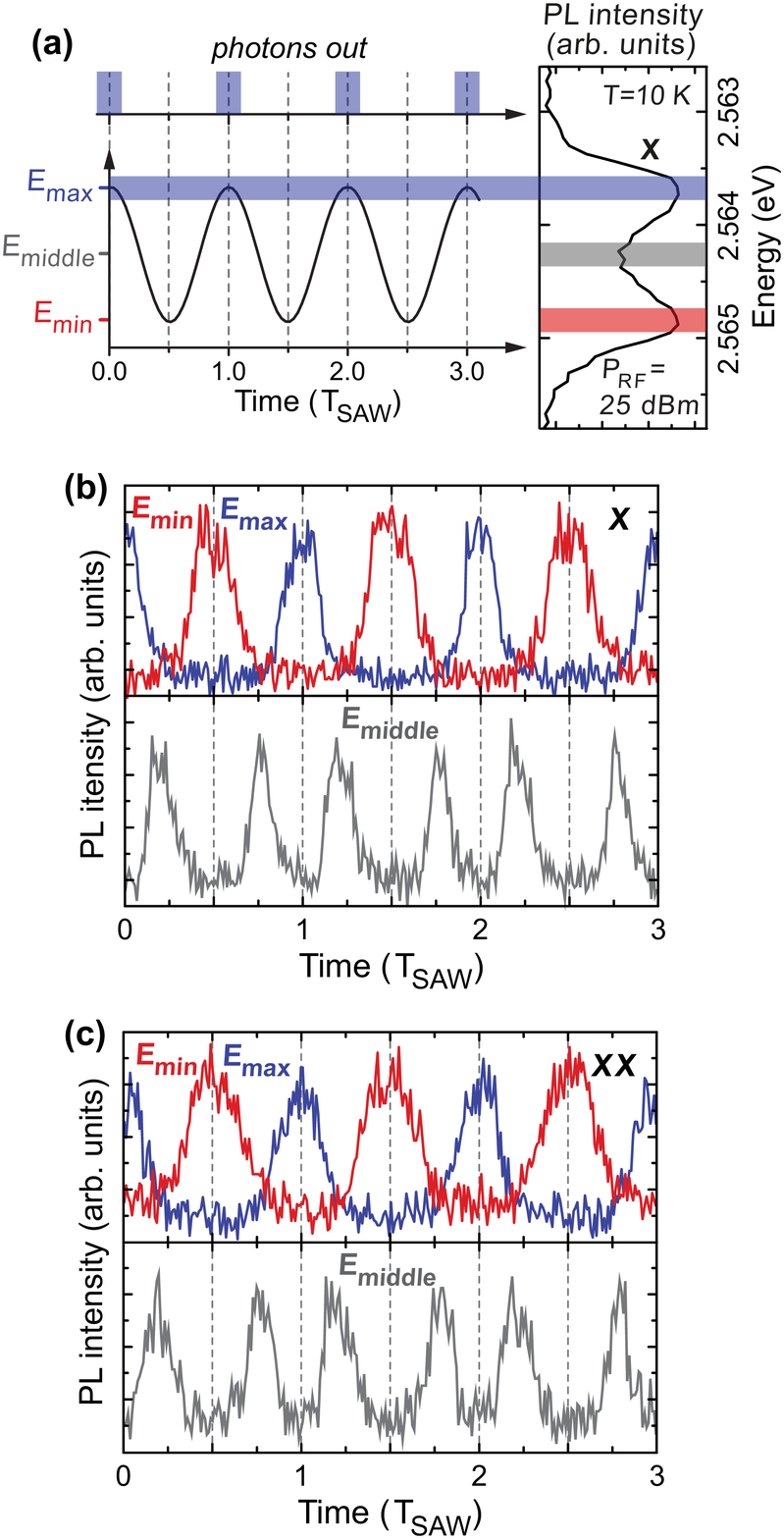}
\caption{.(a) Left panel: Schematics of the SAW-driven spectral modulation of the QD emission energy as a function of time. Colored rectangles labeled $E_{min}$ (red), $E_{middle}$ (grey) and $E_{max}$ (blue) represent the spectral detection window. The temporal emission profile expected for detection at $E_{max}$ is superimposed in the upper part of the panel. Right panel: Time-integrated PL spectrum showing the SAW-split X emission line at $\Prf$=25~dBm. (b-c) Time-resolved PL intensity of the (b) exciton (X) and (c) biexciton (XX) recorded under stroboscopic detection at spectral positions marked by the corresponding colored rectangles in (a).}
\label{fig_S3}
\end{center}
\end{figure}

Figure \ref{fig_S3} displays time-resolved stroboscopic $\mu$-PL experiments synchronized with the frequency of the acoustic wave. They were performed to study the temporal evolution of light intensity emitted at different wavelengths within the SAW-split X and XX PL lines. The center and width of the spectral detection window were determined by the spectrometer grating and slits. The latter were adjusted so as to collect $\sim{20}\%$ of the peak splitting in the time-averaged PL spectra recorded on the CCD camera. The light filtered at the particular wavelength (i.e. $E_{min}$, $E_{middle}$ and $E_{max}$) was then detected using one of the APDs from the HBT setup. The TCSPC electronics was started on a signal derived from the rf generator and stopped on a photon detection by the APD. 

By collecting the photons in a narrow spectral range, the SAW-driven sinusoidal oscillations of the QD emission energy was employed to control the photon emission times. Namely, the SAW forces the QD line to move in and out the selected energy window and, consequently, causes changes in the photon count rate detected at the APD within each SAW cycle. As seen for both X and XX in Fig.~\ref{fig_S3}(b) and Fig.~\ref{fig_S3}(c), respectively, light detected at the highest ($E_{max}$) and lowest ($E_{min}$) energy of the SAW-induced spectral modulation is out of phase with each other and is emitted at SAW frequency. In contrast, at central wavelength ($E_{middle}$) the emission takes place at twice the acoustic frequency, since the QD line moves past this energy twice per SAW cycle. 

As previously reported for III-As QDs~\cite{Gell_APL93_081115_08}, the spectral detection filtering of the SAW-driven dynamic tuning of the QD emission energy enables the photon output to be clocked at the acoustic frequencies. Such acoustically-regulated temporal control of photon emission provides an efficient mechanism for deterministic on-demand generation of single photons without the need for a pulsed laser. By increasing the SAW frequency, high repetition rate (in the GHz range) of the emitted photons can be achieved. 

\subsection{SAW induced emission of anti-bunched photons in GaAs/(In,Ga)As nanowires}
\label{GaAs}

In addition to the local modulation of the photoemission properties, the electric potential of a piezoelectric SAW can also transport electronic carriers in low dimensional semiconductor heterostructures~\cite{Rocke97a, PVS152}. The modulation of the electronic bandstructure by the piezoelectric potential spatially confines electrons and holes at separate positions along the SAW wavelength and transports them with the well-defined velocity of the SAW. This transport mechanism has already been used in planar heterostructures for e.g. the acoustic pumping of electrons and holes from a QW into QD-like recombination centers embedded in quantum wires~\cite{PVS218}. The acoustic pumping mechanism can inject only one electron and one hole per SAW period and, therefore, anti-bunched photons are emitted with a repetition rate matching the SAW frequency. 

\begin{figure}
\begin{center}
\includegraphics[width=\linewidth]{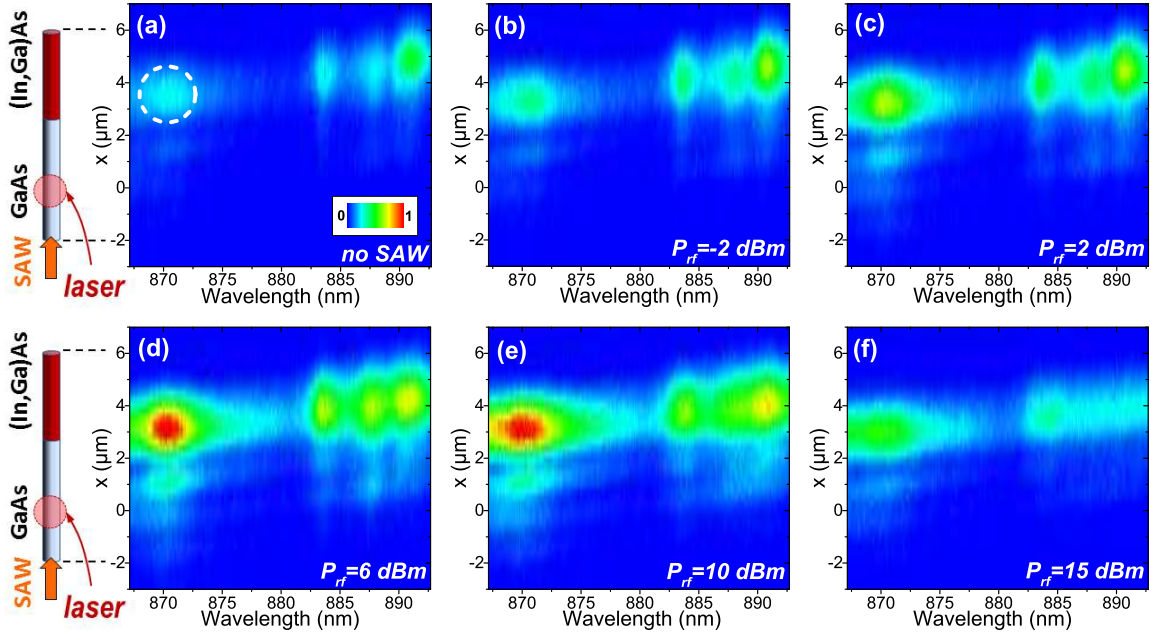}
\caption{Spatially (y-axis) resolved $\mu$-PL spectra of a single GaAs/(In,Ga)As NW under SAWs. The electronic carriers were excited by a laser spot focused on the GaAs segment and acoustically transported towards the (In,Ga)As region, where they recombine at QD-like emission centers (see diagrams on the left side). The PL image was recorded (a) without applied SAW, and (b-f) with increasing SAW powers. The dashed circle in panel (a) shows the emission line probed in the time-resolved PL and autocorrelation measurements of Figs.~\ref{fig_InGaAsTRPL} and \ref{fig_g2vsPrf}, respectively.}
\label{fig_PLvsPrf}
\end{center}
\end{figure}

The acoustic pumping of electronic carriers can also be applied to QD-like centers embedded in semiconductor nanowires~\cite{PVS246, PVS273}. Here, we discuss an example realized in a single GaAs NW containing (In,Ga)As recombination centers at one of its ends~\cite{PVS246}. Figure~\ref{fig_PLvsPrf} displays a series of spatially (vertical scale) and spectrally (horizontal scale) resolved PL maps of the NW at a nominal temperature of 20~K. The electrons and holes were optically excited by a 4~$\mu$W laser beam tightly focused at the NW region whose core consists of GaAs, i.e., $x=0$ at the left inset. Taking into account that the diameter of the NW core is much smaller than that of the laser spot, we estimate that the number of optically generated electron-hole pairs is less than 2000 per laser pulse.

Figure~\ref{fig_PLvsPrf}(a) shows a narrow spectral region between 870-890~nm including a series of spectral lines emitted by the (In,Ga)As segment in the absence of SAWs. The weak remote PL for $x>3~\mu$m is attributed to electron-hole pairs that are either directly generated at the (In,Ga)As segment by the tail of the laser spot, or arrive to it just after laser excitation by carrier diffusion along the NW. When a SAW with frequency $f_3=338$~MHz is applied, the intensity of these emission lines initially increases with the power of the rf signal applied to the IDT [Figs.~\ref{fig_PLvsPrf}(b)-(d)]. The origin of this PL enhancement is the acoustic transport of electrons and holes from the GaAs towards the (In,Ga)As segment. For high rf-powers ($\Prf>10$~dBm), the strong piezoelectric field also modulates the energy levels of the recombination centers, thus leading to \textit{(i)} a broadening of the emission lines [compare the shape of the emission center at 870~nm in panels (d) and (e)], and \textit{(ii)} carrier extraction from the QD-like centers in addition to carrier injection, and therefore to the quenching of the SAW-induced PL signal emitted by the (In,Ga)As region, cf. Fig.~\ref{fig_PLvsPrf}(f).

Figure~\ref{fig_PLvsPrf} also shows weak lines within the GaAs segment with the same energy as the (In,Ga)As emission lines. The intensity and spectral width of these lines follow closely the ones irradiated from the (In,Ga)As segment for all acoustic powers. We attribute these lines to the guiding of the light generated at the (In,Ga)As segment through the GaAs region, which is transparent to the photon energies emitted by the recombination centers in the (In,Ga)As segment. 

\begin{figure}
\begin{center}
\includegraphics[width=.7\linewidth]{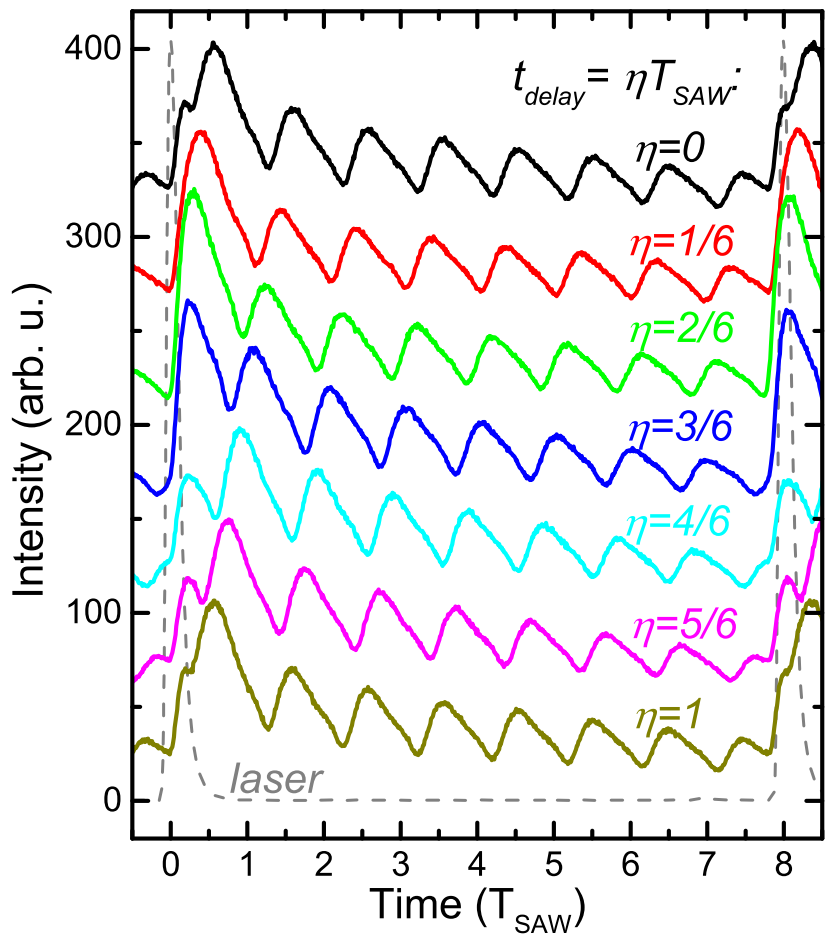}
\caption{Time-resolved $\mu$-PL of the 870~nm emission line of Fig.~\ref{fig_PLvsPrf} (marked with a dashed circle) under a SAW frequency of $\fSAW=338$~MHz and period $\TSAW=2.96$~ns, measured at different time delays, $t_\mathrm{delay}$, between laser pulse and SAW phase. The dashed grey line shows the time-resolved shape of the laser pulse.}
\label{fig_InGaAsTRPL}
\end{center}
\end{figure}

Among the emission lines displayed in Fig.~\ref{fig_PLvsPrf}, the one at 870~nm [marked with a dashed circle in panel (a)] shows the strongest dependence on acoustic power. In addition, this center is also spectrally isolated from the others, thus making its selection easier using band pass filters. We have studied the emission properties of this line by means of time-resolved PL. For this purpose, the pulsed laser was triggered by a reference frequency, $f_\mathrm{laser}$, fulfilling the condition $f_\mathrm{laser}=\fSAW/8$, and the emission intensity was recorded by an APD after rejection of the laser by a long pass filter and of other emission lines of the NW by a band pass filter centered at 870~nm with a FWHM of 10~nm. The curves in Fig.~\ref{fig_InGaAsTRPL} represent time-resolved traces of the PL collected by our detection setup under the same excitation conditions as in Fig.~\ref{fig_PLvsPrf}(d). Each trace corresponds to a different delay between laser pulse and SAW phase, which was settled by electrically delaying the laser trigger by $t_{delay}$. For all traces, the photon emission takes place in the form of pulses whose repetition period corresponds to that of the acoustic wave, $\TSAW=2.96$~ns. Similar SAW-induced spatio-temporal dynamics of photogenerated electrons and holes was also recently observed in GaAs/AlGaAs core/shell NWs of polytipic type \cite{Kinzel2016}. The number of emission pulses between two consecutive laser excitations in Fig.~\ref{fig_InGaAsTRPL} agrees with our laser/SAW synchronization condition.  This proves the role of the SAW in the transport of electronic charges towards the remote recombination centers. 

In our experiment, $\lSAW=11.67~\mu$m, while the transport distance between the generation and the recombination point is barely 4~$\mu$m. As discussed in Ref.~\cite{PVS246}, under these conditions the SAW-induced motion of the electronic carries in the NW becomes similar to the slide of a ball on an oscillating seesaw, with electrons and holes moving in opposite directions for a fixed SAW phase. In this way, the SAW transports electrons to the recombination center in one half cycle and holes in the subsequent half cycle. The persistence of the recombination pulses over times exceeding the laser repetition period, $T_{laser}=8\TSAW=23.68$~ns, is attributed to charge storage in the high density of trapping centers due to stacking faults and twins usually present in GaAs NWs~\cite{Kinzel2016, Jahn_PRB85_45323_12, Graham_PRB87_125304_13}. These traps reduce the carrier mobility and thus hamper the acoustic transport efficiency.

\begin{figure}
\begin{center}
\includegraphics[width=.7\linewidth]{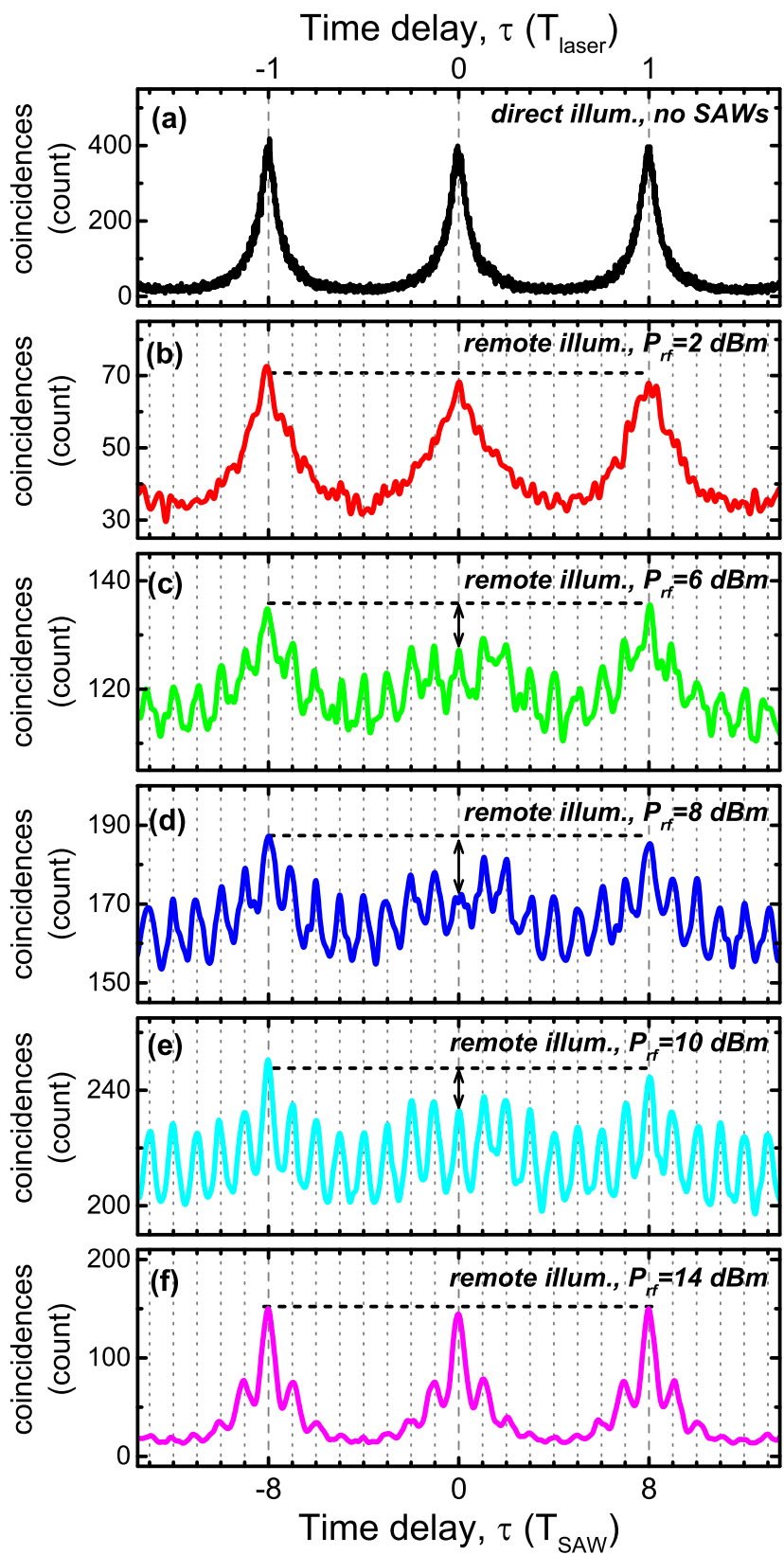}
\caption{Autocorrelation histograms of the 870~nm emission line in Fig.~\ref{fig_PLvsPrf}. (a) Measured under direct illumination of the (In,Ga)As segment of the NW without SAWs excitation. (b-f) Histograms recorded under laser excitation at the GaAs region, see left inset at Fig.~\ref{fig_PLvsPrf}, for increasing rf-powers driving the SAW, $\Prf$.}
\label{fig_g2vsPrf}
\end{center}
\end{figure}

To study the quantum properties of the (In,Ga)As recombination centers, we have measured the histogram of coincidence events [which is proportional to the second-order correlation function $g^{(2)}(\tau)$], of the photons emitted by the 870~nm line. To do this, we have replaced the APD by an HBT setup after the band pass filter in our optical detection system. First, we have analyzed the photon statistics when the (In,Ga)As segment is directly excited by the laser spot in the absence of SAWs. The corresponding histogram of coincidences is displayed in Fig.~\ref{fig_g2vsPrf}(a). It shows a series of peaks with the periodicity of the laser, $T_{laser}$ (only three of these peaks are shown in the figure).  This is the typical shape expected for $g^{(2)}(\tau)$ under pulsed laser excitation \cite{Lounis_RoPiP68_1129_05}. According to quantum theory of light, the value of $g^{(2)}(0)$ is related to the number of photons simultaneously arriving to the HBT detection system, $n$, by the expression \cite{Fox_QO_160_06}:

\begin{equation}\label{Eq_g2}
g^{(2)}(0)=1-1/n.
\end{equation}

\noindent We can estimate the value of $g^{(2)}(0)$ from our histogram of coincidences as the ratio:

\begin{equation}
g^{(2)}(0)=\frac{N(0)-N_b}{N(T_{laser})-N_b}.
\end{equation}

\noindent Here, $N(0)$ and $N(T_{laser})$ are the number of coincidence events measured at $\tau=0$ and $|\tau|=T_{laser}$, respectively, while $N_b$ is the background number of coincidences detected by the HBT system. In the case of Fig.~\ref{fig_g2vsPrf}(a), $g^{(2)}(\tau)\approx 1$, which means that the emission line at 870~nm does not naturally behave as a source of single photons. We attribute this to the fact that the (In,Ga)As segment contains many quantum centers emitting simultaneously within the spectral window allowed by our band pass filter.

Figures~\ref{fig_g2vsPrf}(b)-(f) display the autocorrelation histograms of the 870~nm line under remote laser excitation at the GaAs NW core (same conditions as Figs.~\ref{fig_PLvsPrf} and \ref{fig_InGaAsTRPL}). For low SAW powers, cf. Fig. \ref{fig_g2vsPrf}(b), the shape of the histogram is similar to the one measured under direct laser excitation of Fig.~\ref{fig_g2vsPrf}(a), but with a lower rate of coincidences. However, the number of photons simultaneously arriving to the HBT just after laser excitation is still large enough to prevent the observation of a clear anti-bunching signature

As we further increase the rf-power applied to the IDT, the SAW-induced injection of electrons and holes towards the (In,Ga)As segment begins to take place. The consequence is the emission of light in the form of the SAW-induced peaks discussed in Fig.~\ref{fig_InGaAsTRPL}, and the enhancement of the time-integrated intensity of the measured photoluminescence lines, cf. Fig.~\ref{fig_PLvsPrf}(d). This is reflected in the autocorrelation histogram as follows: \textit{(i)} an increase of the overall rate of coincidences with respect to lower SAW powers, and \textit{(ii)} a change in the shape of the histogram, with the apparition of a series of new peaks fulfilling the SAW/laser synchronization condition $T_{laser}=8\TSAW$. As in Fig.~\ref{fig_InGaAsTRPL}, this is an evidence that the transport of the electronic carriers along the NW and their recombination at the (In,Ga)As region is mediated by the SAW piezoelectric field. More important, for the histograms measured under $P_{rf}=6$, 8 and 10~dBm, $N(0)$ is clearly lower than $N(T_{laser})$, which means that $g^{(2)}(0)<0$. This is characteristic of photon anti-bunching and is a proof of the non-classical behavior of the emission centers under remote acoustic pumping of the electronic carriers.

Finally, at very large SAW powers, the strong piezoelectric field induces not only carrier injection into the QD-like centers, but also carrier extraction, thus quenching the SAW-induced photon bursts. As a consequence, the amplitude of the SAW-related peaks in the autocorrelation histogram reduces, and the intensity of the rate of coincidences decreases, cf. Fig.~\ref{fig_g2vsPrf}(f), in agreement with the intensity decrease of the time-averaged photoluminescence lines discussed in Fig.~\ref{fig_PLvsPrf}(f). Therefore, the anti-bunching signature disappears because the photon statistics is not dominated any more by the SAW-induced, not-classical photon emission. 

\begin{figure}
\begin{center}
\includegraphics[width=.7\linewidth]{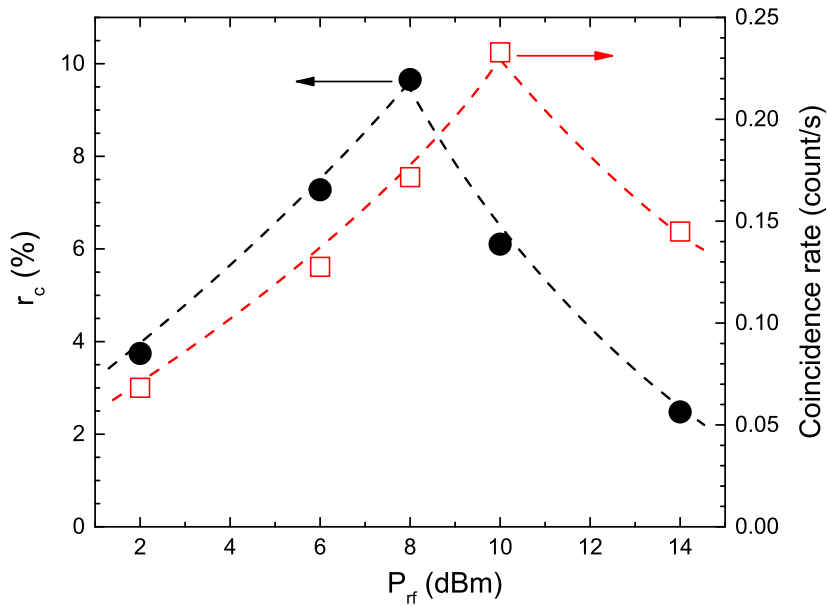}
\caption{Dependence of the degree of anti-bunching, $r_c$ (solid circles), and the average coincidence rate (open squares), on the acoustic power, $\Prf$. The dashed lines are a guide to the eye.}
\label{fig_rcvsPrf}
\end{center}
\end{figure}

We have summarized these observations in Fig.~\ref{fig_rcvsPrf}, where we compare the dependence on acoustic power of the rate of coincidences at $|\tau|=T_{laser}$ (red open squares), with the degree of anti-bunching, $r_c=1-g^{(2)}(0)$ (black solid circles), observed in Fig.~\ref{fig_g2vsPrf}. Although the maxima of the two curves are slightly shifted, it is clear that high anti-bunching levels correlate well with an efficient SAW-induced transfer of the photogenerated electrons and holes from the excitation point towards the recombination centers.
 
To conclude this section, from the maximum measured degree of anti-bunching in Fig.~\ref{fig_rcvsPrf}, $r_c=0.1$, and applying Eq.~\ref{Eq_g2} into our definition of $r_c$, we estimate that no less than $n=1/r_c\approx 10$ photons per SAW period are simultaneously present in our detection system.  Although this number is still too high for applications in single-photon sources, it can be reduced by diminishing the size of the collection area or the bandwidth of the spectral detection. The most promising approach, however, would certainly be the optimization of the growth conditions to create one single (In,Ga)As QD-like center at a well-defined position within the GaAs NW.

\section{Conclusion}
In summary, we have demonstrated the SAW-driven acousto-mechanical modulation of non-classical light emitted from a single GaN/(In,Ga)N nanowire QD. In addition to the fine-tuning of the excitonic transition energies (up to 1.5~meV) by the propagating SAWs, we have employed the spectral filtering combined with the stroboscopic detection technique to control the timing of the photon output. In this way, the SAW-driven dynamic spectral tuning reported here can be readily used to control the QD emission times and obtain triggered single photon sources operating at acoustic frequencies.

The piezoelectric potential of the SAW can also be used to transport photo-excited electrons and holes along a single GaAs nanowire and to inject carriers into an ensemble of (In,Ga)As recombination centers that are spatially separated from the excitation area. The carrier population and depopulation mechanism induced by the acoustic modulation limits the number of recombination events taking place simultaneously in the ensemble, thus allowing the emission of anti-bunched photon pulses with the repetition frequency of the surface acoustic wave.

\ack

The authors acknowledge Prof. Dr. J. M. Calleja, Dr. H. van der Meulen and Dr. P. Corfdir for discussions.  We thank Dr. \v{Z}. Ga\v{c}evi\'c and Prof. Dr. E. Calleja for the fabrication of the GaN/(In,Ga)N nanowire samples, and Dr. C. Somaschini and Dr. L. Geelhaar for the fabrication of the GaAs/(In,Ga)As nanowires. We also thank S. Krau\ss, W. Seidel and A. Tahraoui for assistance in the preparation of the SAW delay lines. S. L. acknowledges financial support from the Spanish MINECO (RyC-2011-09528).

\section*{References}

\bibliographystyle{unsrt}

\end{document}